\documentclass[aps,prl,twocolumn,superscriptaddress,nopacs,a4paper]{revtex4-1}

\usepackage{graphicx}

\usepackage{color}

\pdfoutput=1 

\usepackage{fancyhdr}
\usepackage[colorlinks=true,citecolor=blue,linkcolor=magenta]{hyperref}
\hypersetup{
pdfauthor = {Patrick Maletinsky},
colorlinks = true, linkcolor = blue, urlcolor=blue, bookmarksnumbered =  true}

\newcommand{\ket}[1]{\left|#1\right>}

\begin{document}
\title{Resolved sidebands in a strain-coupled hybrid spin-oscillator system}
\author{J. Teissier}
\affiliation{Department of Physics, University of Basel, Klingelbergstrasse 82, Basel CH-4056, Switzerland}
\author{A. Barfuss}
\affiliation{Department of Physics, University of Basel, Klingelbergstrasse 82, Basel CH-4056, Switzerland}
\author{P. Appel}
\affiliation{Department of Physics, University of Basel, Klingelbergstrasse 82, Basel CH-4056, Switzerland}
\author{E. Neu}
\affiliation{Department of Physics, University of Basel, Klingelbergstrasse 82, Basel CH-4056, Switzerland}
\author{P. Maletinsky}
\affiliation{Department of Physics, University of Basel, Klingelbergstrasse 82, Basel CH-4056, Switzerland}
\email[]{patrick.maletinsky@unibas.ch}

\date{\today}

\begin{abstract}
We report on single electronic spins coupled to the motion of mechanical resonators by a novel mechanism based on crystal strain. Our device consists of single-crystalline diamond cantilevers with embedded Nitrogen-Vacancy center spins. Using optically detected electron spin resonance, we determine the unknown spin-strain coupling constants and demonstrate that our system resides well within the resolved sideband regime. We realize coupling strengths exceeding ten MHz under mechanical driving and show that our system has the potential to reach strong coupling. Our novel hybrid system forms a resource for future experiments on spin-based cantilever cooling and coherent spin-oscillator coupling.
\end{abstract}

\maketitle

Recent years have brought significant advances in the control of nanoscale mechanical oscillators, which culminated in 
experiments to prepare such oscillators close to their quantum ground state\,\cite{Chan2011,Teufel2011} or a single-phonon excited state\,\cite{OConnell2010}.
Generating and studying such states and further extending quantum control of macroscopic mechanical oscillators brings exciting perspectives for high precision sensing, quantum technologies\,\cite{Rabl2009} and fundamental studies of the quantum-to-classical crossover\,\cite{Treutlein2012,Schwab2005,Marshall2003}. An attractive route towards these goals is to couple individual quantum two-level systems to mechanical oscillators and thereby enable efficient oscillator cooling\,\cite{Wilson-Rae2004} or state transfer\,\cite{Rabl2010a} between quantum system and oscillator in analogy to established concepts in ion trapping\,\cite{Leibfried2003}. 
A prerequisite for most of these schemes\,\cite{Wilson-Rae2004,Rabl2010b,Leibfried2003} is the resolved sideband regime, where the transition between the two quantum states exhibits well-resolved, frequency-modulated sidebands at the oscillator Eigenfrequency. 
Various hybrid systems are currently being explored in this context and include mechanical oscillators coupled to cold atoms\,\cite{Camerer2011}, superconducting qubits\,\cite{OConnell2010}, quantum dots\,\cite{Bennett2010,Yeo2013} or solid state spin systems\,\cite{Arcizet2011,Kolkowitz2012}. None of these systems however has reached the resolved sideband regime thus far and novel approaches are needed to further advance quantum control of macroscopic mechanical systems. 

An important aspect that distinguishes existing hybrid systems is the physical mechanism they exploit to couple the quantum system to the oscillator. 
Coupling through electric\,\cite{Bennett2010} or magnetic\,\cite{Arcizet2011,Kolkowitz2012} fields, through optical forces\,\cite{Camerer2011} and strain-fields\,\cite{Yeo2013} have been demonstrated as of now.
Such strain-coupling is based on electronic level-shifts\,\cite{Wilson-Rae2004,Maze2011} induced by crystalline strain during mechanical motion. This type of coupling is particularly appealing in the context of hybrid systems for several reasons: 
On one hand, strain coupling has been predicted to result in interesting and unique system-dynamics, such as strain-induced spin-squeezing \,\cite{Bennett2013} or phonon-lasing\,\cite{Kepesidis2013} and can be used for coherent, mechanical spin-driving\,\cite{MacQuarrie2013}.
On the other hand, strain coupling brings decisive technological advantages as it is intrinsic to the system. It thereby allows for monolithic and compact devices which are robust against manufacturing errors or thermal drifts and thus particularly amenable for future low-temperature operation. 
Despite these attractive perspectives, only few studies have exploited strain-coupling for hybrid systems up to now\,\cite{Yeo2013,MacQuarrie2013}
and strain-coupling of single spins to nanomechanical oscillators has not been demonstrated in any system thus far. 

\begin{figure}
\includegraphics{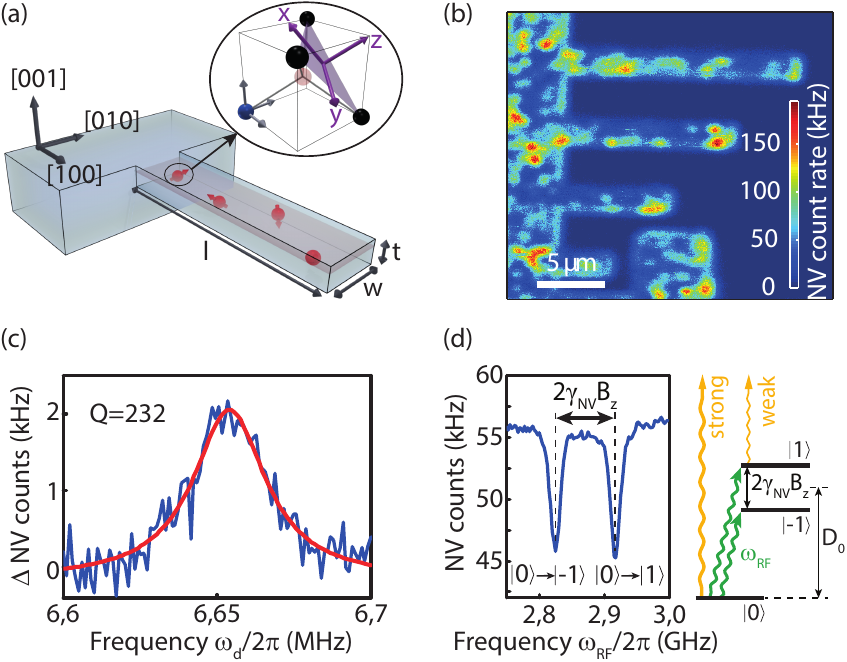}
\caption{\label{FigOverview} 
a.) Schematics of the hybrid device studied in this work. Isolated electronic spins (red arrows) in form of negatively charged Nitrogen Vacancy (NV) centers in diamond (inset) are embedded in a mechanical oscillator and coupled to the oscillators flexural motion through crystal strain. 
b.) Confocal image of our cantilever devices showing individual, implanted NV centers scattered across the sample surface.
c.) Mechanical excitation spectrum of a selected cantilever with resonance frequency and quality factor typical for the devices employed here. The mechanical resonance was measured by monitoring the drop in NV fluorescence from the cantilever end, as $\omega_d$ was varied across $\omega_{\rm mech}$\,\cite{SOM}. 
d.) Optically detected electron spin resonance (ESR) of an NV center in a cantilever. The image to the right illustrates the NVs electronic ground state spin configuration, which consists of a spin-triplet, whose sub-levels can be optically read out, since the $\ket{\pm1}$ states yield less NV fluorescence than the $\ket{0}$ state.}
\end{figure}

In this Letter, we demonstrate for the first time the coupling of a mechanical resonator to an embedded single spin through lattice strain and present clear spectroscopic evidence that our system resides well within the resolved sideband regime.
The devices we study consist of a single crystal diamond cantilever with micron-scale dimensions, which contains an embedded single spin in form of a negatively charged Nitrogen-Vacancy (NV) center [Fig.\,\ref{FigOverview}]. We achieve resolved sideband operation owing to the high mechanical frequency and sizeable coupling strength in our structure. In addition, our room-temperature experiments yield the first quantitative determination of the previously unknown spin-strain coupling constants for the NV ground state.

Our diamond cantilevers consist of single crystalline, ultra-pure diamond (Element Six) and were fabricated through recently developed top-down diamond nanofabrication techniques\,\cite{Maletinsky2012,Ovartchaiyapong2012}. We fabricated our structures on the surface of $\left(001\right)$-oriented diamond starting material and aligned our cantilevers to within few degrees to the $\left[100\right]$ direction of the diamond [Fig.\,\ref{FigOverview}(a)]. Cantilever dimensions were in the range of $10$-$50\times3.5\times0.2$-$1~\mu$m$^3$ for length $l$, width $w$ and thickness $t$, respectively. The corresponding resonance frequencies $\omega_{\rm mech}$ of the fundamental flexural mode of our cantilevers are estimated from Euler-Bernoulli thin beam theory and lie in a range of $1-10~$MHz. A typical mechanical excitation spectrum\,\cite{SOM} is shown in Fig.\,\ref{FigOverview}(c) and yields $\omega_{\rm mech}=2\pi\times6.659~$MHz, a linewidth of $\Delta\omega_{\rm mech}=2\pi\times28.7~$kHz and a resulting quality factor $Q=\omega_{\rm mech}/\Delta\omega_{\rm mech}=232$. The relatively modest value of Q is caused by clamping losses and our experimental conditions under atmospheric pressure, but does not pose a limitation to the experiments discussed here. 

The NV centers in our cantilevers were created through $^{14}$N ion-implantation and subsequent sample annealing in vacuum at $800~^\circ$C. We chose an implantation density that allowed us to isolate single NV centers and an implantation energy resulting in an approximate depth of our NV centers of $10-15~$nm from the cantilever surface. This depth forms a good compromise between NV spin coherence, which improves as depth is increased\,\cite{OforiOkai2012}, and NV-cantilever strain coupling, which increases with proximity of the NV to the surface\,\cite{Wilson-Rae2004}. We address the NV centers in our cantilever devices through a confocal microscope and show a typical confocal fluorescence image in Fig.\,\ref{FigOverview}(b). NV centers are clearly visible as bright spots scattered throughout the device at a density that allows us to  address individual NVs. We drive and detect NV spin-transitions through a nearby microwave antenna and well-established\,\cite{Gruber1997} optical NV spin readout to perform optically detected electron spin resonance (ESR) [Fig.\,\ref{FigOverview}(d)]. 

\begin{figure}
\includegraphics{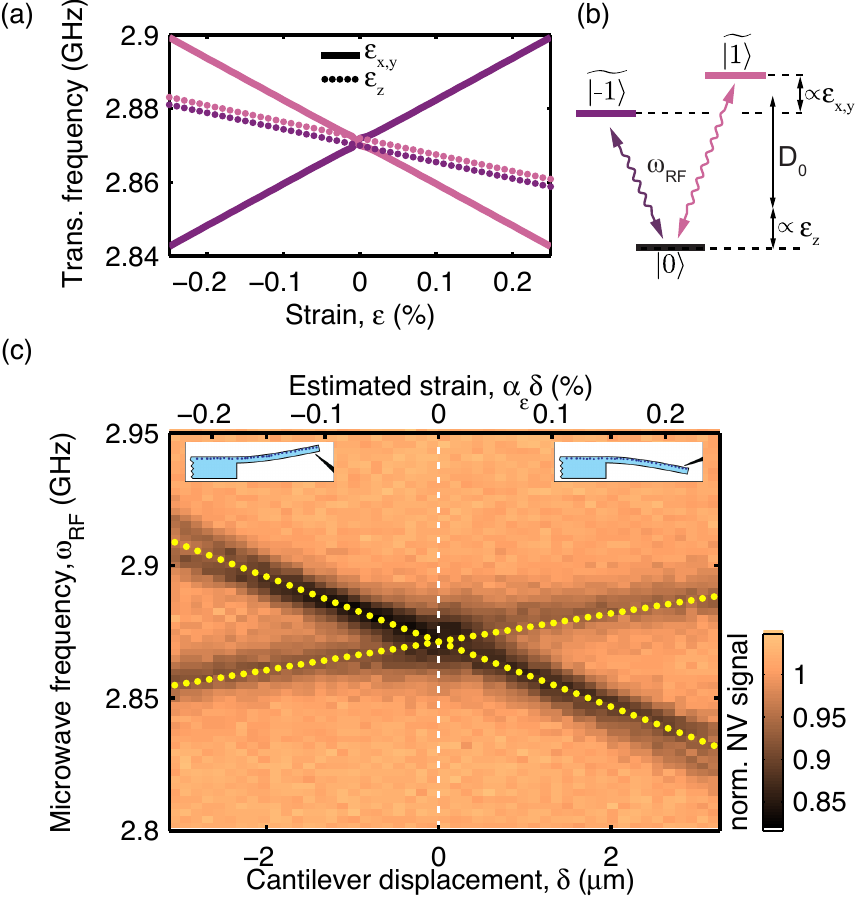}
\caption{\label{FigBending} 
a.) Calculated shifts of NV ESR transition frequencies as a function of crystal strain. The effects of strain transverse ($\epsilon_{x,y}$) and longitudinal ($\epsilon_z$) to the NV axis are plotted separately (solid and dotted lines, respectively).
b.) Schematic of the NV electronic spin states and the action of $H_{\rm strain}$. $\epsilon_{x,y}$ splits $\widetilde{\ket{-1}}$ and $\widetilde{\ket{1}}$, while $\epsilon_z$ shifts $\widetilde{\ket{\pm1}}$ with respect to $\ket{0}$.
c.) Strain-splitting of NV ESR lines as a function of static cantilever-displacement. Positive and negative values of $\delta$ correspond to two different data-sets (separated by the vertical dotted line) and represent tensile and compressive strain at the NV location, respectively (see inset). 
}
\end{figure}

In the following, we will focus on strain-coupling of the NV's ground state electronic spin sub-levels to the motion of our diamond cantilevers. The NV ground state consists of a spin S=1 system with $S_z$ Eigenstates $\left\{\ket{-1},\ket{0},\ket{1}\right\}$ and a zero-field splitting of $D_0=2.87~$GHz between $\ket{0}$ and $\ket{\pm1}$ [Fig.\,\ref{FigOverview}(d)]. The degeneracy of $\ket{\pm1}$ is lifted if the NV experiences magnetic or strain-fields and the ground state spin-manifold can be described by the Hamiltonian\,\cite{Maze2011,Dolde2011}
\begin{equation}
\label{EqnNVHGS}
H/h = \underbrace{\vphantom{_\parallel}D_0 S_z^2}_{H_0}+ \underbrace{\vphantom{_\parallel}\gamma_{\rm\scriptscriptstyle NV}  \vec S \cdot \vec B}_{H_{\rm Zeeman}} + \underbrace{\tilde{d}_\parallel \epsilon_z S_z^2  - \tilde{d}_\perp/2 \left[\epsilon_+ S_+^2 + \epsilon_- S_-^2\right]}_{H_{\rm strain}},
\end{equation}
where $\gamma_{\rm\scriptscriptstyle NV}=2.799~$MHz/G, $h$, $\vec B$ and $\vec S$ are the NV gyromagnetic ratio, Plancks  constant, the external magnetic field and the NV electron spin-operator, respectively. $S_x$, $S_y$ and $S_z$ denote the components of $\vec S$ and $S_+$ and $S_-$ are the spin raising and lowering operators. 
$H_{\rm strain}$ describes the coupling of the NV spin to lattice strain 
$\epsilon_i$ along coordinate $i$ ($i\in \{ x,y, z\}$ as defined in Fig.\,\ref{FigOverview}(a)). $\tilde{d}_\parallel$ and $\tilde{d}_\perp$ are the coupling constants corresponding to strain longitudinal and transverse to the NV axis and $\epsilon_\pm=-\epsilon_y\mp i \epsilon_x$. 
The effect of $\epsilon_z$ is equivalent to a modification of $D_0$\,\cite{Fang2013} and therefore only affects the energy-difference between the states $\ket{\pm1}$ and $\ket{0}$ without mixing any of the zero-field Eigenstates. 
Conversely, a non-zero $\epsilon_{x,y}$ mixes $\ket{1}$ and $\ket{-1}$, which evolve into new Eigenstates $\widetilde{\ket{1}}$ and $\widetilde{\ket{-1}}$. In the limit of strong strain ($\tilde{d}_\perp\epsilon_{x,y}\gg\gamma_{\rm\scriptscriptstyle NV} |\vec B|$) the energy difference between $\widetilde{\ket{\pm1}}$ increases linearly with $\epsilon_{x,y}$ and $\widetilde{\ket{\pm1}}=(\ket{1}\pm\ket{-1})/\sqrt{2}$ [Fig.\,\ref{FigBending}(a) and\,\ref{FigBending}(b)].

In order to experimentally determine the unknown coupling constants $\tilde{d}_\parallel$ and $\tilde{d}_\perp$, we applied variable degrees of strain to an NV center close to the clamping-point of a cantilever by controlled cantilever bending. To that end, we employed a tungsten tip (Omniprobe, Autoprobe 250) mounted on a piezoelectric actuator and positioned this tip on the non-clamped end of the cantilever (with $t=1~\mu$m and $l=45~\mu$m). We then displaced the tip to statically bend the cantilever, which in turn induced compressive or tensile lattice-strain at the site of the NV
\footnote{Note that the spring-constant of the manipulator was orders of magnitude higher than for the diamond cantilever, allowing us to directly relate the applied piezo-displacement to the induced cantilever bending-amplitude}.
We measured the effect of this strain on the NV by monitoring the optically detected ESR spectrum of the NV center as a function of the  displacement $\delta$ of the cantilever's free end. 
Fig.\,\ref{FigBending}(c) shows the result of this experiment: As expected, the zero-field ESR line splits with cantilever displacement as a result of transverse strain. Additionally, a weak center-of-mass shift of the two resulting ESR lines is caused by $\epsilon_z$. We fitted the observed ESR line shifts by diagonalising Hamiltonian\,(\ref{EqnNVHGS}) (white dashed lines in Fig.\,\ref{FigBending}(c)), and obtained strain-coupling constants $\tilde{d}_\parallel=5.46\pm0.31~$GHz and $\tilde{d}_\perp=19.63\pm0.40~$GHz, 
where errors denote $95~$\% confidence intervals of our fits. For the fit, we assumed that the induced strain-field at the cantilever surface $\epsilon_{\left[100\right]}$ is unidirectional and points along the direction of the cantilever (i.e. $\left[100\right]$), as expected from Euler-Bernoulli thin beam theory. Near the clamping-point of the cantilever, we then find $\epsilon_{\left[100\right]}=\frac{3}{2} \frac{t}{l^2} \delta=\alpha_{\epsilon}^{\left[100\right]} \delta$, which for our cantilever yields $\alpha_{\epsilon}^{\left[100\right]}=7\times10^{-4}~\mu$m$^{-1}$. Within this approach, the ratio of $\epsilon_{z}$ to $\epsilon_{x,y}$ is constant and given by the orientation of the cantilever with respect to the NV axis, which in our case [Fig.\,\ref{FigOverview}(a)] yields $\epsilon_z=\sqrt{\frac{2}{3}}\epsilon_{\left[100\right]}$ and $\epsilon_{x,y}=\sqrt{\frac{1}{3}}\epsilon_{\left[100\right]}$ for all NVs. 

Our determination of $\tilde{d}_\parallel$ and $\tilde{d}_\perp$ is qualitatively consistent with theoretical expectations\,\cite{SOM} and yields similar values on most NVs we studied. Interestingly, we also observed NVs whose ESR spectra showed significantly different behaviour in response to beam-bending, compared to the NV presented in Fig.\,\ref{FigBending}\,\cite{SOM}.
Both, the magnitude of the measured strain-shift per cantilever displacement and the ratio of transverse to longitudinal strain varied by up to one order of magnitude in some cases. While we expect $\tilde{d}_{\parallel,\perp}$ to be constant for all NVs, we assign these observations to variations in direction and magnitude of the local strain-field at different NV sites for a given cantilever displacement. In particular close to surfaces, strain-fields are known to exhibit strong variations on the nanoscale\,\cite{Beche2013} due to crystal imperfections and boundary effects. It is thus plausible that for some NVs the local strain-field deviates strongly from expectations based on Euler-Bernoulli theory. 

\begin{figure}
\includegraphics{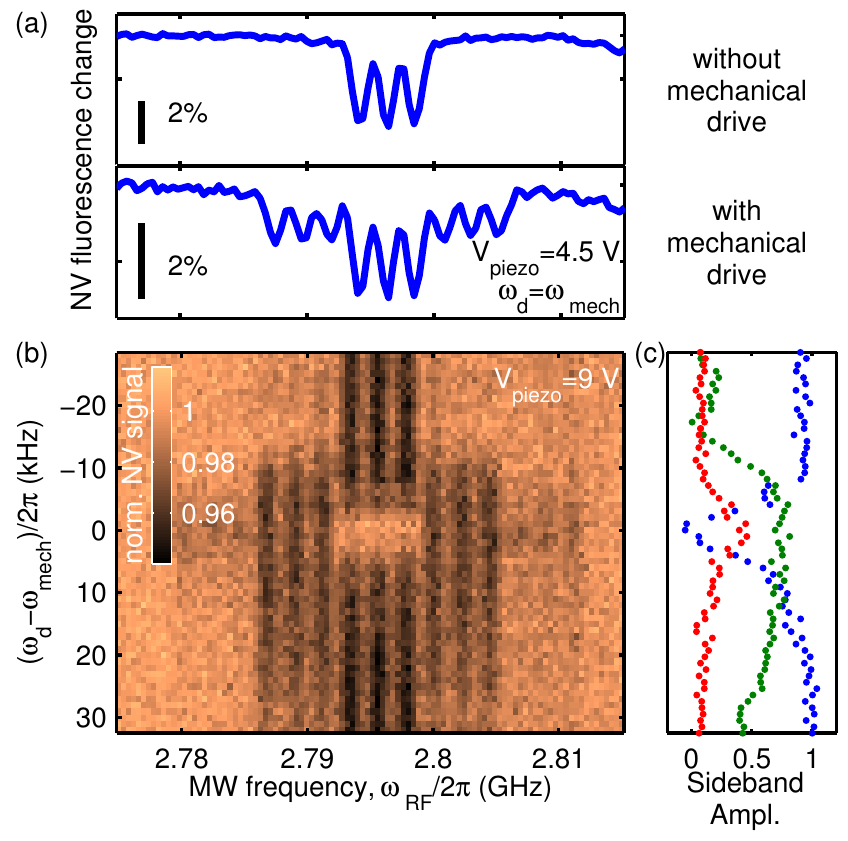}
\caption{\label{FigSidebands} 
a.) ESR trace of the NV in the cantilever in the absence (upper trace) and presence (lower trace) of resonant mechanical excitation. Without excitation, the ESR line exhibits a splitting into three hyperfine components. Mechanical driving induces sidebands to this central carrier at frequencies $\omega_0\pm\omega_{\rm mech}$, where $\omega_0=2.796~$GHz is the bare ESR frequency of the NV center.
b.) Evolution of the ESR sidebands as a function of mechanical drive frequency $\omega_d$ at an excitation amplitude $V_{\rm piezo}=9~$V. A clear resonant behaviour is observed with a maximal sideband amplitude appearing when $\omega_d=\omega_{\rm mech}$.
c.) Amplitude of the carrier signal (blue) and the two sidebands ($n=1,2$ in green and red, respectively) as determined by a Lorentzian fit to the observed ESR dips. A slight asymmetry in the sideband amplitudes with respect to $\omega_d$ is caused by the onset of a mechanical nonlinearity of the diamond mechanical oscillator.}
\end{figure}

After we have established a significant coupling of NV spins to cantilever bending through static strain, we now turn our attention to the dynamics of our hybrid spin-oscillator system. To that end, we provided a mechanical drive to the diamond cantilever by means of a piezoelectric transducer, placed in proximity to our sample and driven at a frequency $\omega_d$ with voltage $V_{\rm piezo}$. We then characterised the resulting dynamical spin-cantilever interaction through high-resolution ESR spectroscopy\,\cite{Dreau2011}. For this experiment, we chose an NV where the induced strain field acts purely longitudinally. Additionally, we applied a magnetic field $B_z=26~$G along the NV axis, such that our discussion can be restricted to the two-level subspace spanned by $\ket{0}$ and $\ket{-1}$ and mixing of $\ket{1}$ and $\ket{-1}$ by transverse strain can be neglected. 
The cantilever drive can then be described by a classical phonon field, which leads to a time-modulated term $\tilde{d}_\parallel \epsilon_z^{\rm max} S_z$ cos$(\omega_{\rm d} t)$\,\cite{Rohr2014} in Hamiltonian\,(\ref{EqnNVHGS})\footnote{We note that in this limit our model is equivalent to spin-oscillator systems coupled through magnetic field gradients\,\cite{Rabl2009, Arcizet2011, Kolkowitz2012}.}, where $\epsilon_z^{\rm max}=\alpha_{\epsilon}^z\delta^{\rm max}$ and $\delta^{\rm max}$ is the maximal cantilever amplitude. 
Using static beam-bending, we measured a strain shift of $\alpha_{\epsilon}^z\tilde{d}_\parallel=19~$MHz per micron of cantilever displacement\,\cite{SOM} for the NV investigated here. The result of dynamic strain-modulation can be seen in Fig.\,\ref{FigSidebands}(a), where we choose $\omega_d=\omega_{\rm mech}$ and compare high resolution NV ESR spectra of the $\ket{0}\rightarrow\ket{-1}$ transition in the presence and absence of the mechanical excitation. Without mechanical drive (upper trace), we observed the well established hyperfine structure of the NV electron spin, which consists of three ESR lines split by the $^{14}$N hyperfine coupling constant $\omega_{\rm HF}=2\pi\times2.166~$MHz\,\cite{He1993}. Upon resonant mechanical excitation however, two clearly resolved, mechanically induced sidebands appear for each of the three hyperfine split ESR lines at detunings $\pm \omega_{\rm mech}$, respectively. This experiment demonstrates that our system resides well within the resolved sideband regime of spin-oscillator coupling, since the ESR line-width $\Delta\omega<\omega_{\rm mech}$ by a factor of three ($\Delta\omega=2\pi\times1.8~$MHz).

To prove the resonant character of our optomechanical coupling and the mechanical origin of the observed sidebands, we extended the experiment presented in Fig.\,\ref{FigSidebands}(a) by sweeping $\omega_d$ over a frequency range of $\pm30~$kHz around $\omega_{\rm mech}$, while monitoring the NVs ESR spectrum [Fig.\,\ref{FigSidebands}(b)]. Clearly, sidebands only appear under resonant driving, when $\omega_d\approx\omega_{\rm mech}$. Furthermore, the frequency range over which sidebands can be observed [Fig.\,\ref{FigSidebands}(c)] closely matches $\Delta\omega_{\rm mech}$ as determined from Fig.\,\ref{FigOverview}(c). This observation demonstrates that the observed sidebands are indeed induced by the mechanical oscillator and in particular excludes sidebands occurring through accidental modulation of the NV spin splitting by electric or magnetic stray fields.

\begin{figure}
\includegraphics{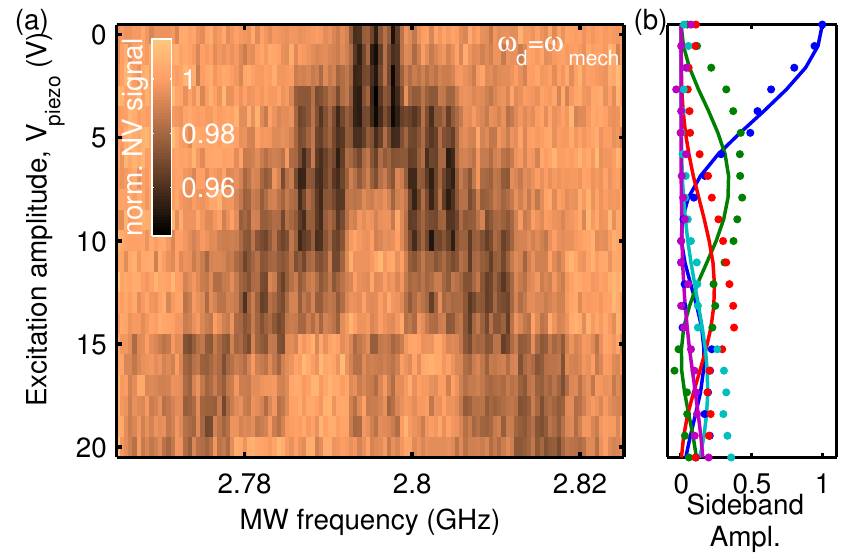}
\caption{\label{FigAmpSw} a.) Amplitude of ESR sidebands for increasing excitation drive-voltage, $V_{\rm piezo}$ of the mechanical oscillator. b.) Relative amplitudes of carrier-signal (blue) and sidebands as a function of excitation amplitude (dots). We determined the amplitudes by  Lorentzian fit to the ESR peaks in a) and normalised the dips in each line by the total measured ESR signal strength. Sidebands of order $n=1,2,3$ are color-coded in red, green and grey, respectively. Lines indicate squares of $n-$th order Bessel functions.}
\end{figure}

Finally, we investigate the evolution of the motion-induced sidebands as a function of the strength of the mechanical drive. Figure\,\ref{FigAmpSw}(a) shows a series of high-resolution NV ESR traces recorded at various strengths of piezo excitation with $\omega_d=\omega_{\rm mech}$. For increasing $V_{\rm piezo}$, we observe an increase of the sideband amplitude and eventually the appearance of higher-order sidebands up to order $n=3$. As expected\,\cite{Leibfried2003}, the amplitude of the $n$-th sideband is well fitted by $J^2_n(\tilde{d}_\parallel \epsilon_z^{\rm max}/\omega_{\rm mech})$ where $J_n(x)$ is the $n$th-order Bessel function of the first kind [see fits in Fig.\,\ref{FigAmpSw}(b)]. Next to a further confirmation of the nature of the sidebands, this measurement allows us to determine the strain coupling constant $\tilde{d}_\parallel$ in this dynamical spin-strain coupling mode. We can extract the modulation depth $m=\tilde{d}_\parallel \epsilon_z^{\rm max}/\omega_{\rm mech}$ as a function of drive amplitude and use an estimated mechanical susceptibility $\chi_{\rm mech}=\delta^{\rm max}/V_{\rm piezo}\approx23~$nm/V\,\cite{SOM} of our system to relate $m$ to $V_{\rm piezo}$. This estimate yields $\alpha_{\epsilon}^z\tilde{d}_\parallel\approx76~$MHz$/\mu$m and lies within a factor of four of our earlier measurement of $\tilde{d}_\parallel$, which is reasonable, given the uncertainty in our estimation of $\chi_{\rm mech}$.

With spin-strain coupling and the resolved sideband regime clearly established,
the question arises to what extent our system is amenable for future experiments in the quantum regime and in particular, whether the strong coupling regime ($g_0^2>\gamma_{\rm mech}\Gamma_{\rm\scriptscriptstyle NV}$) can be achieved. The single phonon coupling strength $g_0=\tilde{d}_{\parallel,\perp} \alpha_{\epsilon} x_{\rm ZPM}$ for our system is determined by the amount of strain per zero point motion $x_{\rm ZPM}=\sqrt{\hbar/(2m_{\rm eff}\omega_{\rm mech})}$ generated at the NV location and is largely determined by cantilever geometry ($g_0\propto\sqrt{1/l^3w}$). 
While for our cantilevers we find $g_0\approx2\pi\times0.25~$Hz an increase of $g_0$ to the kilohertz range\,\cite{Bennett2013} is within reach by further reducing dimensions of the mechanical oscillator\,\cite{Babinec2010}. 
$g_0$ has to be benchmarked against both the NV spin dephasing rate $\Gamma_{\rm\scriptscriptstyle NV}=2\pi/T_2$ (with $T_2$ the NV spin coherence time) as well as the oscillators thermal decoherence rate $\gamma_{\rm mech}=k_B T/\hbar Q$ (with $k_B=210~$GHz/K the Boltzmann constant and T the bath temperature). 
With NV $T_2$ times approaching one second at low temperatures\,\cite{Bar-Gill2013} and a projected value of $\gamma_{\rm mech}=2\pi\times1~$kHz for $T=100~$mK and $Q=10^6$\,\cite{Ovartchaiyapong2012}, the strong coupling regime thus appears realisable in our system. The variations of strain coupling constants we observed additionally suggests that strain engineering in our devices could be employed to increase even $g_0$ further.

In summary, we have established NV centers embedded in single crystalline diamond nanomechanical resonators as a valuable resource for future experiments with hybrid systems in the quantum regime. In particular, we have firmly established the resolved sideband regime and quantitatively determined the NV-oscillator coupling strength. 
The non-trivial form of the spin-strain coupling Hamiltonian opens opportunities for exploring highly interesting avenues such as spin induced oscillator sideband cooling\,\cite{Wilson-Rae2004}, spin squeezing\,\cite{Bennett2013} or ultrafast, mechanical spin driving\,\cite{MacQuarrie2013}. 
Finally, strain coupling of orbitally excited states is five orders of magnitude stronger\,\cite{Batalov2009,Togan2011} compared to the values we established, which would bring our system deep into the ultrastrong coupling regime ($g_0\gg\omega_{\rm mech}$). Extending our experiments to cryogenic operation where coherent coupling to these states become accessible thus forms another highly exciting perspective.

We thank A.\,H\"ogele. P. Treutlein, S.D.\,Huber and M. Pickova for fruitful discussions and valuable input. We gratefully acknowledge financial support through the NCCR QSIT, a competence center funded by the Swiss NSF, through the Swiss Nanoscience Institute and through SNF Grant No. 200021\_143697/1.

\bibliographystyle{apsrev4-1}
\bibliography{BibNVcantilever}

\newpage

\section*{Supplementary Information}

\section*{S1. Estimation of $\chi_{\rm mech}$ and determination of mechanical resonance frequency}

\begin{figure}
\includegraphics{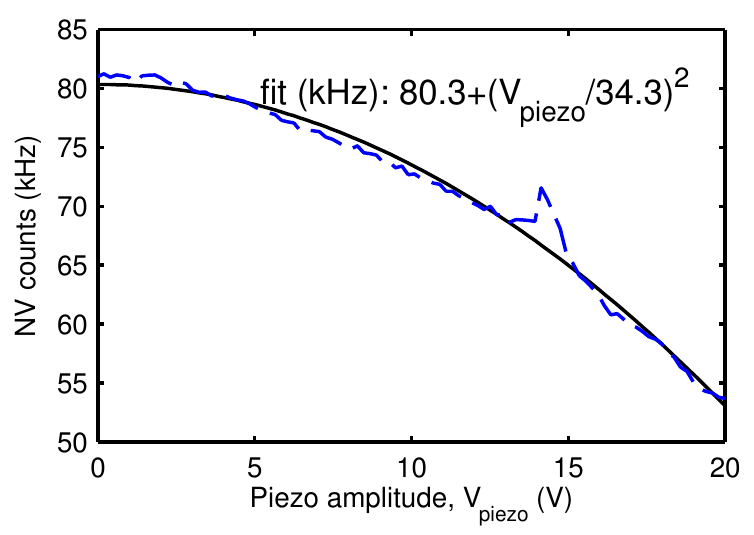}
\caption{\label{FigAmpCal} 
Variation of NV count rate collected from the cantilever's free end as a function of cantilever piezo excitation amplitude. The measured curve can be used to estimate the mechanical susceptibility, $\chi_{\rm mech}$ of our cantilever system (see text).}
\end{figure}

We estimated the mechanical susceptibility, $\chi_{\rm mech}$, of our cantilevers, i.e. the excitation amplitude, $a^{\rm max}$ as a function of the piezo drive voltage, $V_{\rm piezo}$, by monitoring the deflection of the non-clamped end of the cantilever as a function of excitation voltage. To that end, we monitored NV fluorescence originating from the cantilever end and recorded the drop in detected NV fluorescence as a function of $V_{\rm piezo}$ (Fig.\,\ref{FigAmpCal}). An approximate calibration of the cantilever excitation amplitude was then possible through the knowledge of the point-spread function (PSF) of our confocal microscope.

In our determination of $\chi_{\rm mech}$, NV fluorescence was collected from the focal spot of our confocal microscope and cantilever oscillation then led to a reduction of the detected count rate as the cantilever end spent less time in the microscope's collection spot. To formalise this situation, we assumed an approximative axial PSF (i.e. the PSF along the optical axis, $z'$) given by a Gaussian $P(z')=\mathrm{exp}(-(z'\sqrt{4{\rm ln}(2)}/\Delta z'_{\rm FWHM})^2)$ with a full-width at half maximum (FWHM) of $\Delta z'_{\rm FWHM}=1.26\lambda/NA^2$\,\cite{Webb1996}, 
where $NA$ is the numerical aperture of the microscope objective and $\lambda$ the wavelength of fluorescence light. For our setting ($\lambda=750~$nm, $NA=0.8$), we obtain $\Delta z'=1477~$nm. 

The time-averaged collected NV fluorescence can then be calculated as $\int_0^{2\pi} P(a^{\rm max}\mathrm{sin}(t)) \mathrm{d}t=2\pi*\mathrm{exp}\left(-2\left(\frac{a^{\rm max}}{\Delta z'}\right)^2\right)I_0\left(2{\rm ln}(2)\left(\frac{a^{\rm max}}{\Delta z'}\right)^2\right)$, with $I_n(x)$ the $n$-th order modified Bessel function of first kind. To second order in $z'$ we then obtain an NV fluorescence rate proportional to $1-2{\rm ln}(2)*\left(\frac{a^{\rm max}}{\Delta z'}\right)^2=1-2{\rm ln}(2)*\left(\frac{\chi_{\rm mech}V_{\rm piezo}}{\Delta z'}\right)^2$.

Fig.\,\ref{FigAmpCal} shows the recorded NV fluorescence as a function of $V_{\rm piezo}$ along with the fit to the quadratic function given above. From the fit parameters, we then find $\chi_{\rm mech}=\Delta z'/(34.3*(2{\rm ln}(2))^2)=23~$nm/V, 
which is the value used in the main text of our paper.

The cantilevers mechanical resonance frequency, $\omega_{\rm mech}$, was determined with the same basic technique as described in the previous paragraphs. However, in order to determine $\omega_{\rm mech}$, we applied mechanical excitation at a constant piezo excitation amplitude $V_{\rm piezo}=2~$V and varied the drive frequency $\omega_d$. The resulting variation of NV fluorescence counts as a function of $\omega_d$ is presented in Fig.~1c of the main text and shows the expected resonant behaviour around $\omega_{\rm mech}$.

\section*{S2. Static beam bending experiments}

\begin{figure*}[th!]
\centerline{\includegraphics{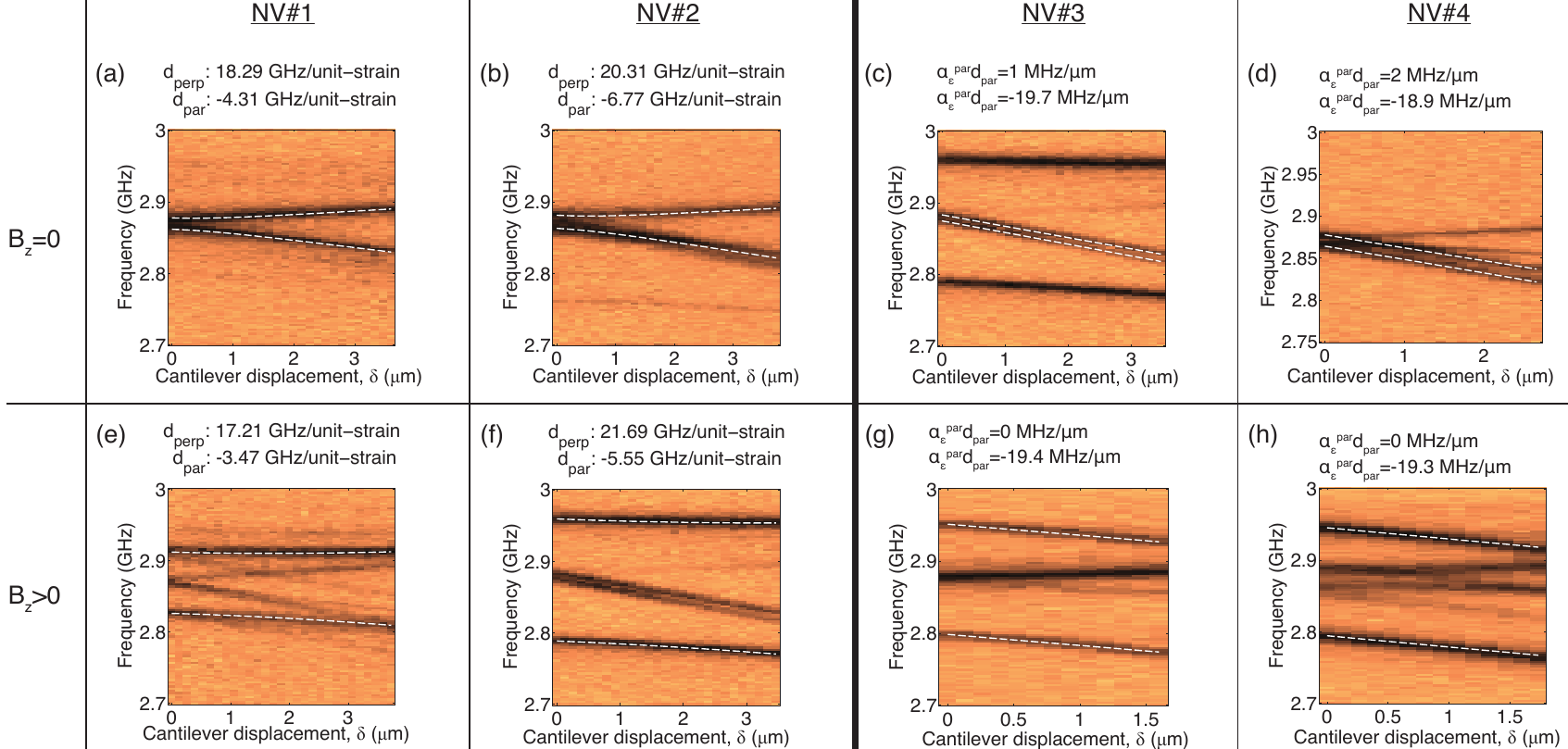}}
\caption{\label{FigMoreBending} 
Additional examples of static beam bending experiments for different NVs. NV\#1 and \#2 show a response to bending as expected from Euler-Bernoulli theory, while NV\#3 and \#4 experience longitudinal strain only. The experiments were conducted as described for Fig.\,(2) of the main text and were performed without ((a)-(d)) and with ((e)-(h)) an external magnetic field applied along the NV axis. The NV response to beam-bending was fitted based on Hamiltonian (1) in the main text (dashed white lines). For NV\#1 and \#2, we used $\tilde{d}_{\parallel}$ and $\tilde{d}_{\perp}$ as fit parameters and quote the resulting numbers in the sub-figures. In the case of NV\#3 and \#4, the experimental data was fit using $\tilde{d}_{\parallel}\alpha_{\epsilon}^{\parallel}$ and $\tilde{d}_{\perp}\alpha_{\epsilon}^{\perp}$ as free parameters (see text). For all fits, error bars were comparable to the values stated in the main text. In (g) and (h), we enforced $\tilde{d}_{\perp}\alpha_{\epsilon}^{\perp}=0$, due to exceedingly large error-bars on this parameter. Note that the some of the data sets contain ESR signatures of two NVs, which however we could clearly distinguish by their different response to magnetic and strain fields.
}
\end{figure*}

Figures\,\ref{FigMoreBending}(a)-(d) show static beam bending experiments we performed on several NV centers on different cantilevers in addition to the data presented in the main text of our paper. We determined the longitudinal and transverse strain-coupling coefficients by fits based on Hamiltonian~(1) and noted the resulting values in the subfigures. For NV\#1 and NV\#2, we find values for $\tilde{d}_{\parallel,\perp}$ which are comparable to the strain-coupling strengths we determined for the NV presented in Fig.\,(2) of the main text. We assign the small scatter in $\tilde{d}_{\parallel,\perp}$ to slight variations in the local strain environments of the NVs.

Additionally, we also present data on NV\#3 and NV\#4, which show significant deviations from the behaviour of NV\#1 and NV\#2 in that there, strain acts only along the NV axis (i.e. the zero-field ESR line only shows a shift, but no splitting). For these NVs, the strain field induced by cantilever bending thus appears to be oriented predominantly along one of the $\left<111\right>$ axes; a situation which Euler-Bernoulli theory fails to describe. In these cases, it makes little sense to use $\tilde{d}_{\parallel,\perp}$ as free fitting parameters and instead, we extend our formalism and define $\alpha_{\epsilon}^{\parallel,\perp}$ such that $\epsilon_{\parallel,\perp}=\alpha_{\epsilon}^{\parallel,\perp} \delta$. We then use $\tilde{d}_{\parallel}\alpha_{\epsilon}^{\parallel}$ and $\tilde{d}_{\perp}\alpha_{\epsilon}^{\perp}$ as free fitting parameters and give the resulting values in Fig.\,\ref{FigMoreBending}(c,d) and (g,h). We stress that $\tilde{d}_{\parallel,\perp}$ should be constant for all NVs and we assign the observed variations of strain-coupling coefficients to deviations of $\alpha_{\epsilon}^{\parallel,\perp}$ from Euler-Bernoulli theory (which would predict $\alpha_{\epsilon}^{\parallel}=\sqrt{\frac{2}{3}}\alpha_{\epsilon}$ and $\alpha_{\epsilon}^{\perp}=\sqrt{\frac{1}{3}}\alpha_{\epsilon}$, as stated in the main text). Note that NV~\#4 was used for our experiments demonstrating resolved sideband operation presented in Fig.(3) and Fig.(4) of the main text.

On top of these beam-bending experiments on additional NVs, we also performed static beam bending in the presence of an external magnetic field $B_z$ along the NV axis. $B_z$ splits $\ket{1}$ and $\ket{-1}$ in energy and suppresses their mixing by $\epsilon_{x,y}$ to first order, so that coupling to longitudinal strain can be measured without any possible influence from transverse strain components. Indeed, this expectation was confirmed by the corresponding beam-bending experiment  (Fig.\,\ref{FigMoreBending}(e)-(h)), which we performed in presence of an external magnetic field ($B_z=16~$G in (e) $B_z=26~$G in (f,g,h)). The longitudinal strain coupling constants we determined in these experiments were largely consistent with the values obtained at $B_z=0$.

\section*{S3. A priori estimates for $\tilde{d}_\parallel$ and $\tilde{d}_\perp$ }

To provide an a priori estimate of the coupling strength of the NV spins to strain, we start by the observation that coupling of NV spins to strain is  equivalent to their coupling to electric fields\,\cite{Maze2011}. We can therefore relate $\tilde{d}_\parallel$ and $\tilde{d}_\perp$ to the recently measured electric-field coupling constants $d_\parallel=0.35~$Hz/(V/cm) and $d_\perp=17~$Hz/(V/cm)\,\cite{Dolde2011} through the piezoelectric constant $g_{\parallel,\perp}=\partial \epsilon_{\parallel,\perp}/\partial E_{\parallel,\perp}$. $g_{\parallel,\perp}$ has never been directly measured, but theoretical estimates yield $g_{\parallel,\perp}\approx2.4\times10^{-10}~$(V/cm)$^{-1}$ (see Appendix D in Ref.\,\cite{Maze2011}). We thus expect strain coupling coefficients $\tilde{d}_\parallel=d_\parallel/g_\parallel\approx1.4~$GHz and $\tilde{d}_\perp=d_\perp/g_\perp\approx70~$GHz per unit strain, respectively. Both values agree to within a factor of four with the values we measured on the majority of NVs in our devices.

\end{document}